# How observers create reality

Brian D. Josephson

Mind–Matter Unification Project, Department of Physics, University of Cambridge, J J Thomson Avenue, Cambridge CB3 0HE, UK.

ABSTRACT

Wheeler proposed that repeated acts of observation give rise to the reality that we observe, but offered no detailed mechanism for this.  Here this creative process is accounted for on the basis of the idea that nature has a deep technological aspect that evolves as a result of selection processes that act upon observers making use of the technologies.  This leads to the conclusion that our universe is the product of agencies that use these evolved technologies to suit particular purposes.

---

INTRODUCTION

The idea that observers create reality is associated with the name of John Archibald Wheeler, who in an article entitled Law Without Law(1) proposed that the laws of nature, instead of being fixed, emerge as a result of past observation processes.  This proposal originated from Bohr's interpretation of quantum reality involving complementarity, according to which in the quantum domain apparatus used to determine the value of a variable does not merely gain knowledge of that variable, but in addition causes that variable to take a definite value, with the value prior to the measurement interaction having been indefinite.  Wheeler then asked the question: if such an elementary quantum process is an act of creation, is an act of creation of any other kind needed to bring into being all that is?  To do this would seem to require some kind of organisation, and Wheeler had little to say as regards how this might happen, suggesting only "beginning with the big bang, the universe expands and cools.  After aeons of dynamic development it gives rise to observership. Acts of observer-participancy — via the mechanism of the delayed choice experiment — in turn give tangible "reality" to the universe, not only now but back to the beginning".

The picture proposed here differs from Wheeler's firstly in that we do not presume the big bang to be the start of everything. Accordingly, preparatory action can take place prior to this in some primordial realm. Crucially, we also allow for the existence of 'developed observers', meaning systems that can achieve more complicated outcomes than just simple measurement, in accord with Barad's hypothesis (2) "knowing is not a play of ideas within the mind of a Cartesian subject that stands outside the physical world the subject seeks to know ... knowing is a physical practice of engagement".  In this connection Barad notes also that "we need not reserve the notion of "measurement" for [inter]actions that we humans find useful in laboratory practices, but can understand it more generally.  Phenomena [associated with measurement processes] are not mere human contrivances manufactured in laboratories. Phenomena are constitutive of reality".

In line with these comments, our assumption will be that activity in the quantum domain resembles in various ways the activity of human beings, this resemblance providing a basis for the possibility of systematic creation.  Accordingly we consider what principles underlie such organised activity.  Human development involves specific kinds of activity where a child acquires, in accord with some overall scheme, a range of progressively advanced skills, a given skill typically building upon previously learned ones (3). The fact that these developments can occur depends on effective strategies, implemented in some way by the nervous system architecture.  These individual strategies, implemented in some way by the nervous system, collectively give rise to individuals possessing typical human capacities and in addition, through the capacities for interaction between individuals originating in such developments, give rise in addition to the achievements of societies.

THE TECHNOLOGICAL ASPECT OF REALITY

Such developments over the passage of time contrast markedly with the behaviour of physical systems in general, and we now consider factors underlying these differences.  In this connection, it is helpful to invoke the concept of technology. While science studies natural phenomena without regard to their application, technology concerns itself with that subset of phenomena that do have application, that is to say behave in a way suited to some purpose, that is to say bringing into existence some specific outcome. In a sense, technology has the capacity to take over physical reality, without in any way conflicting with the dictates of physics, technological schemes such as the overall developmental scheme involved in human development being particularly effective in its ability to govern physical reality.

A specific human capacity, for example the use of language, involves an overall scheme that may be said to embody a formative logic involving a descriptive narrative that in principle explains how the capacity works, just as mechanisms in general, such chiming clocks or computer software, embody a specific logic that explains how they work. The testing by users of particular components of that logic, being tools brought out to deal with particular situations, plays an important role in determining the successful functioning of the entire system, ensuring for example that the parts work together successfully. For example in the case of language, speakers try out particular tools to address particular kinds of situation, while listeners try to find tools that will have the outcome of a successful response by the listener to the speaker's words. Speaker and listener work together towards situations where their joint use of tools has satisfactory outcomes. The apparent miracle of language working as well as it does depends on the lesser miracle of the existence of tools that address particular problems, and can work well together, and the kinds of product that tools are capable of producing.

The above account has characterised what might be called the technological realm of physics, explained for the case of human development. As defined, biological organisms equally have a technological aspect, involving the numerous systems studied by biologists. In either case the technology was, as it were, hidden beneath the surface, unknown until uncovered by the activities of scientists. The technological realm can provide an answer to Wheeler's problem of how individual creative acts are organised to give rise to the specifics of 'all that there is'. That organisation is simply the organisation of an evolving technology, the components coordinated with each other in accord with the principles discussed.

We can see this through a comparison of human development and the hypothetical development that would solve Wheeler's problem. Humans develop in accord with a scheme oriented towards the development of the type of human behaviour observed, which behaviour is consistent with the need to preserve humans. Similarly the deeper level system discussed would develop in ways that use tools appropriate to some grander scheme that can ensures the system's survival over an extended, possibly infinite duration. The ability to create universes might be an aspect of this overall scheme. In conventional biology reproduction requires chemical entities such as DNA, but technology in general can use alternative copying mechanisms, including ones of a linguistic character.

CIRCULAR THEORY AND THE BASIS PROBLEM

There remains what might be called the basis problem. Human mental capacities are based upon the architecture of the nervous system, while the technologies employed by life are based largely on chemistry. On what would the more fundamental analogue hypothesised here be based? A possible answer is provided by Ilexa Yardley's 'everything's a circle' (4). This approach postulates a basic dynamic involving some kind of circular movement, units of this kind combining to create more complicated dynamics. Such a situation can exhibit the most fundamental aspects of life, survival and reproduction. Structures are then presumed to evolve tools capable of maintaining the bases of these circular movements, while maintaining evolutionary capacities. One possible mechanism would involve the formation of subcommunities developing their own specialisation, as happens with human communities. What exactly would be needed for a system able to sustain its existence for eternity must remain a topic for future investigation.

NATURE BOTH ORDERED AND CHAOTIC: A POST-MATHEMATICAL SCIENCE

This picture is radically different from the kind of theory adopted in modern physics, with its mathematical equations asserting that two quantities are identical. This approach has been remarkably successful in physics, enabling extremely precise predictions of natural phenomena to be made. But this is in part a consequence of selectivity in investigation, since clearly such precision is not possible in all situations, for example those where there is sensitive dependence on initial conditions (chaos).  This matters because there are situations where regularities coexist with chaos.

It may then be the case, as discussed in (5), that in averaging in order to obtain precise predictions, relevant knowledge relating to details of individual systems is lost.  In biology, variability is an ever-present feature, and precise calculations have a limited part to play.  Biological regularities may arise from mechanisms discussed above in terms of the technological aspect of reality: biological systems develop tools and more general schemes which do not necessarily have to fit to some precise model, but merely need to be fit for some purpose, and radically different solutions to the same problem may be possible.  But in some contexts precision is important; some situations demand accurate behaviour while in others (e.g. the process of walking) flexibility may be at least as important.  Conventional physics has taken it as axiomatic that nature can be reduced to a mathematical formula, but it is unclear that nature actually conforms to such a straitjacket.  The picture proposed here, taking into account the important role that the technological aspect of nature can play in determining the nature of phenomena, is less constrained and may be closer to the reality.

Precision in behaviour is not necessarily the same thing as behaviour constrained by well-defined mathematical laws, and the question remains as to why do we find parts of nature that behave in ways that can be characterised precisely in mathematical terms?  A general mechanism that can lead to such behaviour would be the imposition upon a system of a consistent set of rules corresponding to the axioms of a mathematical system.  A practical example is the production of spherical mirrors through grinding together mirror and tool, the sphere being the only shape that is invariant under the transformations of a particular symmetry group, these transformations being enacted physically through the rotations of the mirror during the grinding process.  Here we see the mathematical property of symmetry being enforced by the grinding process, which implants into the resulting physical system mathematically characterisable features.

If the picture developed above is correct, the processes involved may visibly manifest themselves in various ways not well understood in current science; for example 'significant form'(6) in art may be correlated with processes involving a coding system used in connection with signs at the fundamental level, while mathematical capacities may also originate in processes at a fundamental level.  And, in this picture, we do not have to reject as inappropriate the idea that the emergence of man through evolution is not purely accidental, in contrast to the views of Monod (7), since achievement of such an outcome could be a goal of the technologists at the subtler level, the various processes involved in the deeper level working towards such a conclusion.

ACKNOWLEDGEMENTS

Development of these ideas has been influenced by the Circular Theory of Ilexa Yardley, to whom I am indebted for correspondence clarifying the theory.  I am indebted also to Alex Hankey for discussions of organisation in biological systems.